\documentclass{article}
\usepackage{authblk}
\usepackage{multirow}
\usepackage{xcolor}
\usepackage{setspace}
\usepackage{amsmath}
\usepackage{mathtools}
\usepackage{bm}
\usepackage{amsthm}
\usepackage{amssymb}
\usepackage{algorithm}
\usepackage{algpseudocode}
\usepackage{graphicx,psfrag,epsf}
\usepackage{enumerate}
\usepackage{subcaption}
\usepackage{caption}
\usepackage{natbib}
\usepackage{tikz}
\usepackage{url} 
\usepackage{appendix}
\usepackage{titlesec}
\usepackage{etoolbox}
\usepackage{graphicx} 
\usepackage{cancel}
\usepackage{breqn}
\usepackage{comment}

\makeatletter
\def\@seccntformat#1{\@ifundefined{#1@cntformat}%
   {\csname the#1\endcsname\quad}
   {\csname #1@cntformat\endcsname}}
\apptocmd\appendix{%
    \newcommand\section@cntformat{\appendixname\ }
    \addtocontents{toc}{\bigskip\noindent\textbf{Appendix Material}\par}
    {}{}}
\makeatother

\newcommand{\blind}{1}

\begin{document}

\def\spacingset#1{\renewcommand{\baselinestretch}%
{#1}\small\normalsize} \spacingset{1}


\if1\blind
{
  \title{\bf PAM-HC: A Bayesian Nonparametric Construction of Hybrid Control for Randomized Clinical Trials Using External Data} 
  \author[1]{Dehua Bi}  
  \author[2]{Tianjian Zhou}
  \author[3]{Wei Zhong}
  \author[1]{Yuan Ji}
  \affil[1]{Department of Public Health Sciences, The University of Chicago, IL}
  \affil[2]{Department of Statistics, Colorado State University, CO}
  \affil[3]{Oncology Biostatistics, Pfizer Inc.}
  \maketitle
} \fi

\if0\blind
{
  \bigskip
  \bigskip
  \bigskip
  \begin{center}
    {\LARGE\bf PAM-HC: A Bayesian Nonparametric Construction of Hybrid Control for Randomized Clinical Trials Using External Data}
\end{center}
  \medskip
} \fi

\bigskip
\begin{abstract}

It is highly desirable to borrow  information from external data to augment a control arm in a randomized clinical trial, especially in settings where the sample size for the control arm is limited. 
However,  a main challenge in borrowing information from external data 
is to accommodate potential heterogeneous subpopulations  across the external and trial data.  
We apply a Bayesian nonparametric model called Plaid Atoms Model (PAM) to identify overlapping and unique subpopulations across datasets, with which we  restrict the information borrowing to the common subpopulations. 
This forms a hybrid control (HC) that leads to more precise estimation of treatment effects 
Simulation studies demonstrate the robustness of the new method, and an application to an Atopic Dermatitis dataset shows improved treatment effect estimation.

\end{abstract}

\noindent%
\vfill
\newpage
\spacingset{1.45} 

\section{Introduction}

Randomized clinical trials (RCTs) are the gold standard to objectively assess the superiority of a new drug over a control. It is widely acknowledged that RCTs with a 1:1 randomization ratio yield the highest statistical power. Nevertheless, enrolling patients under such a design can sometimes be challenging like in rare diseases, pediatric trials, or settings where an $r$:1 ($r > $1) randomization ratio is used to enhance patient enrollment.
With the availability of historical trial data or real-world data (RWD) like the electronic health records, statistical models have been proposed to borrow information in these data to estimate treatment effects more accurately. For example, when a standard of care has been widely tested or administered in a patient population, the available response data could be used to augment the control arm in a clinical trial and form a hybrid control (HC). Due to the augmented information in the HC, a more precise estimation of the treatment effect could be achieved. 

In drug development, information borrowing from external data for an RCT is regulated. The U.S. FDA recently  has released  a guidance document on the design and conduct of external controlled trials for Drug and Biological products \citep{us2023considerations}. The document emphasizes the importance of ensuring that the trial eligibility criteria can be applied to the external control arm in order to obtain a population comparable to that of the clinical trial. Thus, it is critical to ensure 
the similarity of patient baseline characteristics between 
the external data and the current RCT. Another issue discussed is the extent to which one may borrow information \citep{chen2020propensity}. Historical trial data and RWD can be larger than the current RCT data, and one must be cautious not to let the borrowed information dominate the results of the current trial. Therefore, oftentimes information from external data 
is 
discounted to avoid overwhelming the statistical inference of the current study. 

In the literature, many statistical methods have been proposed to borrow information from external data for RCTs. 
Bayesian models like the power prior (PP) \citep{ibrahim2000power}, commensurate prior (CP) \citep{hobbs2012commensurate}, robust meta-analytic-predictive prior (RMAP) \citep{schmidli2014robust}, and the latent exchangeability prior (LEAP) \citep{alt2023leap} all construct hierarchical models for external and current trial data. Specifically, the PP method assumes that the treatment outcome parameters are the same between the current trial and the external data. It utilizes a discounting factor to reflect the user's prior belief regarding the similarity between the historical data and the current trial, thereby discounting the likelihood of the historical data. CP uses different parameters for the current trial and the external data, but assuming the parameters of the trial follow a prior distribution with a mean equal to the parameters for the external data
. RMAP employs a mixture of a meta-analytic-predictive (MAP) prior, which is an informative prior, and a vague prior (robust component) to mitigate the potential issue of over-borrowing. However, these three methods do not consider situations in which only a subset of patients in the external data are comparable to the current study. 
Recently, \cite{alt2023leap} propose the LEAP prior, which dynamically borrows information from historical trials assuming a subset of individuals in the historical data are exchangeable with the current study. 

Another class of methods utilizes propensity scores (PS) to identify matched patients between the external data and the current trial data. For example, \cite{chen2020propensity} proposed the propensity-score integrated composite likelihood (PSCL) method to address the situation where only a subset of patients in the external data are comparable to the current trial. However, as noted by \cite{chandra2021bayesian}, \cite{king2019propensity}, and \cite{zhao2004using}, matching patients based on their PS does not necessarily imply matching of covariates. In addition, PS-based methods are often sensitive to model specification for estimating the PS.

A recent study by \cite{chandra2021bayesian} introduces a third class of methods that utilizes Bayesian nonparametric models (BNP) to identify ``common clusters" of patients across the current trial and the external data. The BNP model has the ability to automatically cluster patients in the current trial and the external data based on baseline covariates. Their method, called CA-PPMx, assumes the external data consists of all the subpopulations that are present in the current trial. 

Motivated by CA-PPMx, we propose a BNP approach, called PAM-HC, for constructing an HC arm for an RCT using external data. Here, PAM refers to a BNP model in \cite{bi2023pam} that generates overlapping clusters. 
Different from \cite{chandra2021bayesian}, we assume that the current RCT and external data may share common subpopulations of patients, while each may consist of unique ones as well. In other words, 
PAM 
can identify common and unique subpopulations across observations arranged in groups. Using PAM, the HC is constructed by only borrowing information from the common subpopulations between the external data and the control in the RCT. 
We employ a power prior to discount the information borrowing. In addition, the BNP models in the proposed PAM-HC method generate random clusters characterized by a posterior distribution. Therefore, the entire statistical inference is model based and variabilities on the clustering and treatment effect estimates are properly accounted for. 

In the subsequent sections, we first review the PAM method in Section \ref{sec:review}. We provide a detailed description to the proposed PAM-HC method in Section \ref{sec:method}. We present the simulation setup and results in Section \ref{sec:simulation} comparing our method to the PSCL method and a baseline method that does not involve information borrowing. In Section \ref{sec:application}, we showcase an application of PAM-HC to real-life trial data.  Finally, we conclude our work in Section \ref{sec:conclusion}.

\section{Review Plaid Atoms Model (PAM)}
\label{sec:review}

We assume that there is an ongoing RCT 
with an $r:1 \, (r > 1)$ 
randomization ratio between the treatment and control arms. 
In addition, assume there exists an external dataset comprising patients with the same disease that have been treated by the same control in the current RCT. For example, the control arm in the RCT and external data could be a standard chemotherapy and the treatment arm in the RCT could be a new immunotherapy. 
We denote the treatment arm of the current RCT as group 1 ($j = 1$), the control arm as group 2 ($j = 2$), and the external data as group 3 ($j = 3$). Assume there are $n_j$ patients in group $j,$ for $j = 1, 2, $ and $3$. Therefore, $N=n_1 + n_2$ is the sample size of the RCT, and $n_1/n_2 \approx r$ due to the $r:1$ randomization ratio.
We denote $\bm{y}_j = \{y_{i,j}\}_{i=1}^{n_j}$ the patients outcome data in group $j$, and 
$\bm{x}_{i,j} = \{x_{i,j,1}, \ldots, x_{i,j,p}\}$ a $p$-dimensional random vector consisting of the patient $i$'s baseline covariates in group $j$.

We assume that the current trial and the external data consist of heterogeneous subpopulations of patients based on their baseline characteristics (covarivates), with patients from the same subpopulation forming a cluster. For the rest of the discussion, we use the terms ``cluster" and ``subpopulation" of patients interchangeable. Lastly, we refer to overlapping subpopoulations of patients that are present in multiple groups as ``common clusters". Conversely, we use the term ``unique cluster" to describe the subpopulation of patients that is only present in one but not other groups.

\begin{figure}[!p]
\begin{center}
\includegraphics[width=0.6\textwidth]{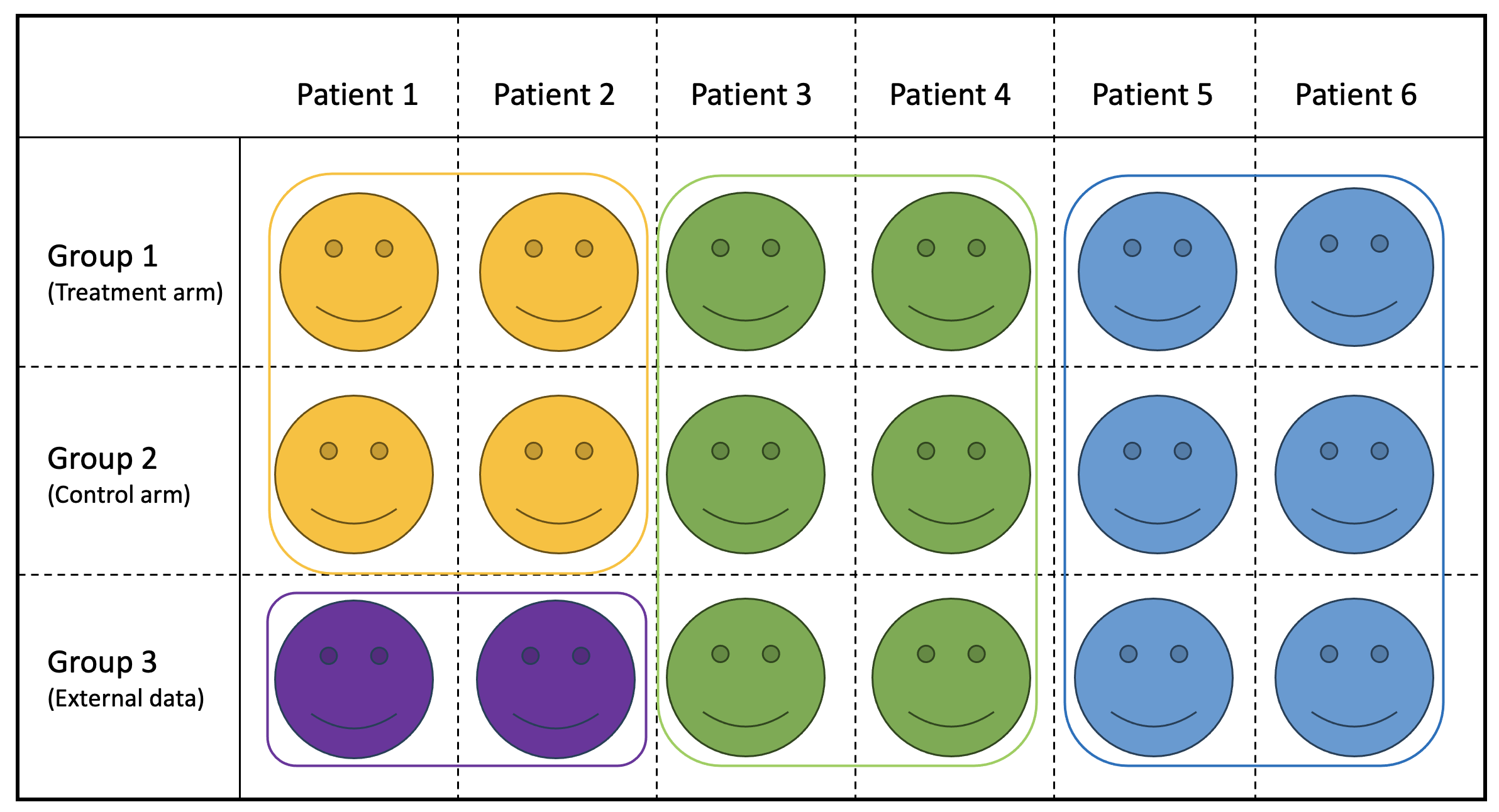}
\end{center}
\caption{An illustration of clustering pattern under PAM. Rows represent groups and columns are patients within each group. The three groups correspond to the current RCT's treatment and control arms, and the external data. There are four homogeneous subpopoulations of patients (clusters) represented by colored smiley faces in blue, green, purple, and yellow. 
The boxes represent the common or unique clusters. For example, the green cluster is common and shared across all three groups, while purple is unique to group 3.}
\label{fig:PAM_pat}
\end{figure}

To find patients clusters in the current RCT and the external data, we adopt PAM in \cite{bi2023pam}. 
An example of the clustering structures identified by PAM is shown in Figure \ref{fig:PAM_pat}.

\noindent In this illustration, each color represents a cluster. 
The blue and green clusters are common and 
the purple cluster 
is unique to group 3. If one knows the clustering pattern in Figure \ref{fig:PAM_pat}, one would only borrow information from the green and blue clusters in the external data because they are shared with the trial data. However, one should not borrow information from the purple cluster since it is unique to the external data. 

A brief review of the statistical model is provided next. For more detail 
refer to  \cite{bi2023pam}. 
Denote $Z_{i,j}$ as the cluster membership indicator for patient $i$ in group $j$, where $\{Z_{i,j} = k\}$ indicates that patient $i$ in group $j$ is assigned to cluster $k$. The Bayesian nonparametrics model in PAM is given by a hierarchical structure as follows: 
\begin{equation} \label{eq:PAM}
\arraycolsep=1.4pt\def\arraystretch{1.25}
\begin{array}{l}
    \bm{x}_{i,j}|Z_{i,j}, \{\bm{\mu}_k, \bm{\Sigma}_k\}_{k=1}^{\infty} \sim MVN\left(\bm{\mu}_{Z_{i,j}},\bm{\Sigma}_{Z_{i,j}}\right) \\
    Z_{i,j}|\{\pi_{j,k}\}_{k=1}^{\infty} \sim \sum_{k=1}^{\infty} \pi_{j,k} \delta_k(Z_{i,j}), \,\, \pi_{j,k} = \pi_{j,k}'\prod_{l=1}^{k-1}(1 - \pi_{j,l}') \\
    f(\pi_{j,k}'|\bm{\beta}, \alpha_0, p_j) = p_j \times f_{\text{Beta}}\left(\alpha_0\beta_k, \alpha_0\left(1 - \sum_{l=1}^k \beta_l\right)\right) + \underbrace{(1 - p_j) \times I(\pi_{j,k}' = 0)}_{(*)} \\
    p_j|a, b \sim Beta(a, b), \,\, \bm{\beta}|\gamma \sim GEM(\gamma), \\
    (\bm{\mu}_k, \bm{\Sigma}_k)|\bm{\mu}_0, \lambda, \bm{\Psi}, \nu \sim NIW(\bm{\mu}_0, \lambda, \bm{\Psi}, \nu),
\end{array}
\end{equation} 
where $MVN$ stands for the multivariate normal distribution, $\delta_A(B)$ is the indicator function ($\delta_A(B) = 1$ if $B \in A$ or $B = A$, and $\delta_A(B) = 0$ otherwise), $f_{\text{Beta}}(a,b)$ is the density function of the Beta(a,b) distribution with mean $a/a+b$, $GEM(\gamma)$ represents 
the Griffths, Engen and McCloskey distribution \citep{pitman2002poisson}, 
distribution $NIW$ stands for to the normal-inverse-Wishart distribution, parameter $\pi_{j,k}$ represents the cluster weights of cluster $k$ in group $j$, and parameters $(\bm{\mu}_k, \bm{\Sigma}_k)$ denote (mean, covaraince matrix) of the $k$th cluster. Additional priors can be assigned to hyperparameters $\gamma$ and $\alpha_0$. 
Model (\ref{eq:PAM}) in PAM largely resembles the well known hierarchical Dirichlet Process (HDP) model \citep{teh2004sharing}, which induces common clusters across groups. PAM adds a unique model component $(*)$, which allows some common cluster to have zero weight in group $j$, thereby producing unique clusters.

Through \eqref{eq:PAM}, PAM generates a joint posterior distribution of all the parameters including the cluster membership 
$\bm{Z} = \{Z_{i,j}\}_{\forall i,j}$. Through $\bm{Z}$ we can easily find the common and unique clusters. Since a posterior distribution of $\bm{Z}$ is generated, the number of clusters and clustering memberships themselves are random. In PAM-HC, we utilize the features of PAM that identifies common and unique clusters, upon which we build models and inference for constructing a hybrid control arm and estimating treatment effects. 

\section{Methodology}
\label{sec:method} 
\subsection{Clustering of patients}
In PAM, 
the cluster membership matrix $\bm{Z} = \{Z_{i,j}\}_{\forall i,j}$ indicates which clusters are common and which are unique. 
Since patients in the same cluster are believed to be ``similar" in their covariates, they are expected to react similarly to the control treatment, under the assumption that the covariates have captured all the factors that are related to treatment response. Of course, when unmeasured confounders are present, the proposed method will be inadequate. Such investigation is beyond the scope of this paper and left for future work. 

\begin{figure}[!p]
\begin{center}
\includegraphics[width=\textwidth]{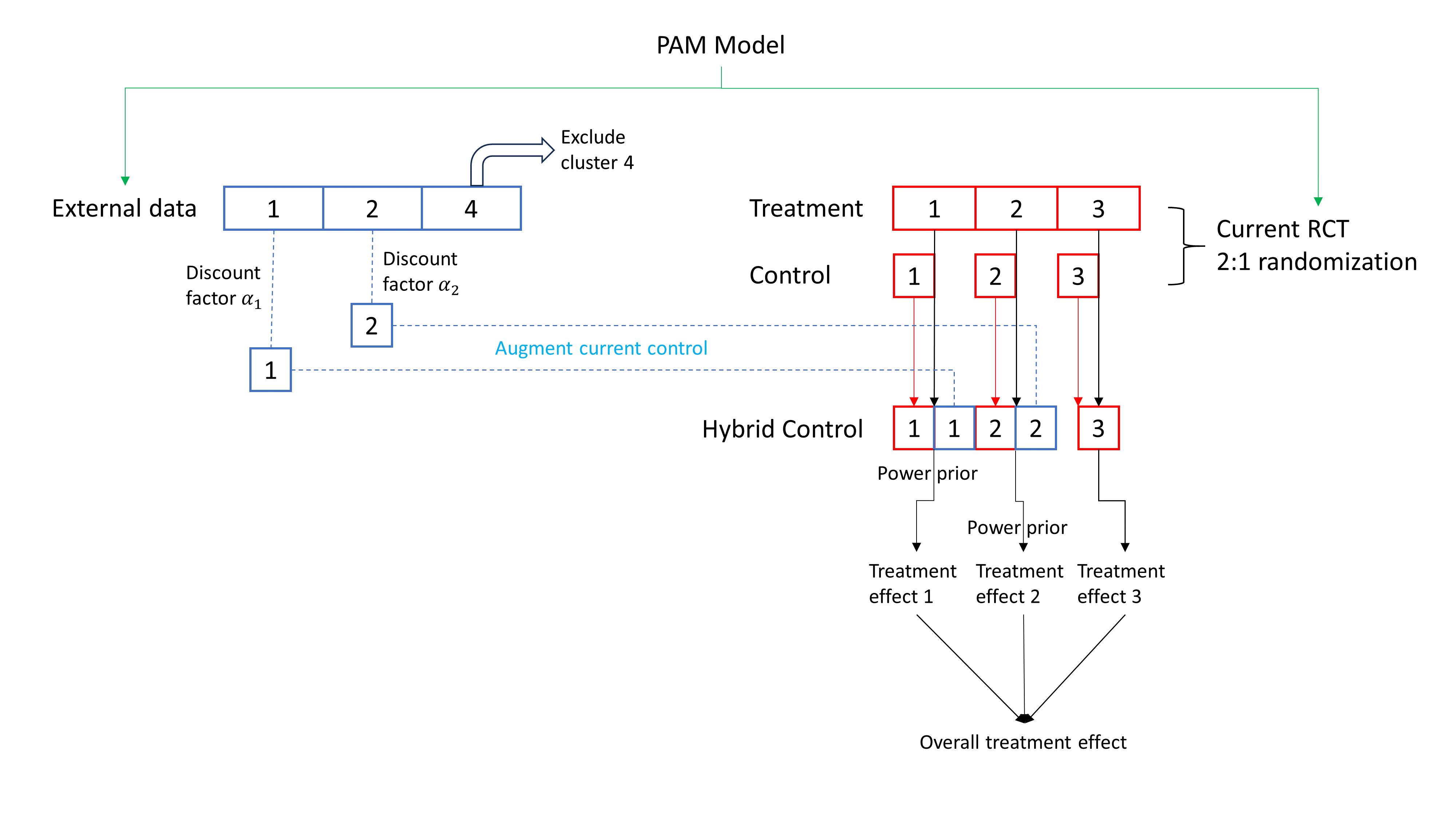}
\end{center}
\caption{A stylized illustration of PAM-HC. Numbers in the boxes denote cluster labels. Boxes in red color represent patients in the RCT and those in blue represent patients in the external data. Cluster 4 is unique to the external data and therefore is not used for forming the HC. Cluster 3 is unique to the RCT and therefore is not augmented.}
\label{fig:PAM_overview}
\end{figure}

Let $A_{j,k} = \{i: Z_{i,j} = k, i = 1, \ldots, n_j\}$ represent the set of patients in the $k$-th cluster in group $j$. 
Also denote the set of cluster labels in each group $j$ as $O_j = \{k: A_{j,k} \neq \emptyset\}$. We define the set of common cluster labels  between a pair of  groups as $C_{j,j'} = \{k: k \in O_j \cap O_{j'} \text{ for } j \neq j', \,\, j,j' \in \{1,2,3\}\}$. Our focus is on set $C_{2,3}$, which consists of the common cluster labels  between the current trial control arm and the external data. 
The proposed PAM-HC method augments $A_{2,k}$ by borrowing information from patients in $A_{3,k}$ for cluster(s) $k \in C_{2,3}$. Figure \ref{fig:PAM_overview} provides a schematic overview of PAM-HC.

\noindent To further illustrate our idea, 
Consider a hypothetical cluster membership $\bm{Z}$ matrix
$$\bm{Z} = \left[\begin{matrix} 1 & 1 & 2 & 2 & 3 & 3 \\
1 & 2 & 3 & & & \\ 1 & 1 & 2 & 2 & 4 & 4 \end{matrix}\right],$$
where the rows are groups and columns are patients. Based on  $\bm{Z}$, there are four clusters, three ($k = 1, 2, 3$) for groups 1 and 2, and three ($k = 1,2,4$) for group 3. 
Clusters 1 and 2 are shared across groups 1 and 2, while cluster 4 is unique for group 3. Also, we have for group 1: $A_{1,1} = \{1,2\}$, $A_{1,2} = \{3,4\}$, $A_{1,3} = \{5,6\}$, and $O_1 = \{1,2,3\}$; for group 2: $A_{2,1} = \{1\}$, $A_{2,2} = \{2\}$, $A_{2,3} = \{3\}$, and $O_2 = \{1,2,3\}$; and for group 3: $A_{3,1} = \{1,2\}$, $A_{3,2} = \{3,4\}$, $A_{3,4} = \{5,6\}$, and $O_3 = \{1,2,4\}$.
The set of common clusters are $C_{1,2} = \{1,2,3\}$, $C_{1,3} = \{1,2\}$, and $C_{2,3} = \{1,2\}$ between the treatment and control arms, between the treatment arm and the external data, and between the control arm and the external data, respectively. We focus on the set $C_{2,3} = \{1,2\}$, and
for clusters $k \in C_{2,3}$, i.e., $k = 1$ and $k = 2$, construct an HC by borrowing information from patients in the external data belonging to clusters 1 and 2, but not cluster 4. 
For illustrative purposes, the stylized example assumes all the clustering memberships are fixed. 
In actual modeling, PAM-HC generates random 
clustering memberships 
which allows for assessment of variabilities in subsequent inference of treatment effects. This will be clear in Section \ref{subsec:inf} later. 

\subsection{Information borrowing across common clusters}
We use the power prior to borrow information across the common clusters between the control and external data in order to form an HC. Similar to \cite{chandra2021bayesian}, our approach involves performing a regression analysis of the outcome variables $y_{i,j}$ on the corresponding covariates $\bm{x}_{i,j}$ through the clusters $A_{j,k}$, $j = 1, 2, 3$. 
Denote $\theta_{1,k}$ and $\theta_{2,k}$ the cluster-specific response parameter in the treatment and control arms, respectively, for cluster $k$. We use a simple hierarchical model for $\theta_{1,k}$ for $k \in O_1$, the clusters in the treatment group. Recall $\bm{y}_1 = \{y_{i,1}\}_{i=1}^{n_1}$ are the observed patient responses in group 1, the treatment group. We assume
$$y_{i,1}|Z_{i,1} = k, \theta_{1,k} \sim F(\theta_{1,k}), $$
$$\theta_{1,k} \sim \pi_0(\theta_{1,k}),$$
where $F(\cdot)$ denotes the likelihood of $y$. For continuous outcome, $F(\theta) = N(\mu,\sigma^2)$, and $\theta_{1,k} = (\mu_{1,k}, \sigma_{1,k}^2)$, and for binary outcome, $F(\theta) = Bern(q)$, with $\theta_{1,k} = q_{1,k}$. In addition, $\pi_0(\theta_{1,k})$ is a vague prior for $\theta_{1,k}$. For $\theta_{2,k}$, the response parameter for group 2, the control arm, we use
$$y_{i,2}|Z_{i,2} = k, \theta_{2,k} \sim F(\theta_{2,k}),$$
and the power prior for $\theta_{2,k}$. 
Recall $\bm{y}_2 = \{y_{i,2}\}_{i=1}^{n_2}$ and $\bm{y}_3 = \{y_{i,3}\}_{i=1}^{n_3}$ are the observed responses in groups 2 (the current control) and 3 (external data), respectively. We assume the prior of $\theta_{2,k}$ is given by 
$$p(\theta_{2,k}|\bm{y}_3, A_{3,k}, \alpha_k) \propto \left[\prod_{i \in A_{3,k}} f(y_{i,3}|\theta_{2,k})\right]^{\alpha_k} \pi_0(\theta_{2,k}) $$
where $f(\cdot|\theta)$ is the p.d.f of $F(\theta)$, $\pi_0(\theta_{2,k})$ is a vague prior for $\theta_{2,k}$, and $\alpha_k \in [0, 1]$ is a discount factor (or power parameter) for cluster $k$. We estimate $\alpha_k$ as a deterministic function of cluster weights $\pi_{j,k}$ and cluster membership $\bm{Z}$.
Specifically, when $k \in O_3$ but $k \notin C_{2,3}$, we set $\alpha_k = 0$. In words, for unique clusters in the external data, there is no borrowing and the power parameter $\alpha_k = 0$. Otherwise, the cluster is shared between the external data and control, and the discount factor $\alpha_k$ is given by \cite{chen2020propensity}:
\begin{equation}\label{eq:ppp}
\alpha_{k} = \min\left(\pi_{2,k}^* \cdot I, \; n_{3,k}\right)/n_{3,k}, \; 
\end{equation}
where $$I=\frac{r-1}{r+1} N, $$
is the total number of patients to be borrowed from external data so that the information in the HC is of the same amount as the treatment arm. The value of $I$ is easily derived based on  the $r:1$ randomization ratio of the RCT and the desired 1:1 matching between the treatment and HC. In addition, 
$n_{3,k} = |A_{3,k}|$, where $|.|$ denotes the cardinality of the set, and 
 $\pi_{2,k}^*$ is the proportion with which we want to borrow from the $I$ patients  for cluster $k$. 
 For example, if $N=300$ and $r=2$, then $I=100$ which means one would borrow information from up to 100 patients from the external data to form an HC so that the amount of information in the HC matches that of information in the treatment arm. 
Within each cluster $k$, we use the following steps to compute $\pi_{2,k}^*$. 
Recall the current trial is randomized with a ratio of $r:1$, $r > 1$, and $\pi_{j,k}$ is the probability (or weights) of cluster $k$ in group $j$ (PAM model \eqref{eq:PAM}). We want to augment the control to form an HC in which the information is worth the same number of patients as the treatment arm in each cluster $k.$ Mathematically, this means 
$$ 
{N \cdot \pi_{1,k}\cdot \frac{r}{r+1}}
= 
{N \cdot \pi_{2,k}\cdot \frac{1}{r+1} + I \cdot \pi_{2,k}^* 
}
, \quad \text{where } I=\frac{r-1}{r+1} N.$$
Solving for $\pi_{2,k}^*$, we have
\begin{equation}\label{eq:pi2ks}
    \pi_{2,k}^* = \frac{1}{r-1}(r\pi_{1,k} - \pi_{2,k}).
\end{equation}
Equation (\ref{eq:pi2ks}) leads to a solution for (\ref{eq:ppp}), and hence a value for $\alpha_k$. In practice, to prevent negative values of $\pi_{2,k}^*$ (when the control arm already has a larger number of patients in cluster $k$ than the treatment arm), we use $\pi_{2,k+}^*$, i.e., $\pi_{2,k+}^* = \pi_{2,k}^*$ if $\pi_{2,k}^* > 0$, and $\pi_{2,k+}^* = 0$ otherwise. 

The construction of $\alpha_k$ in \eqref{eq:ppp} and $\pi_{2,k}^*$ in \eqref{eq:pi2ks} adaptively borrows more or less information for cluster $k$ based on the imbalance in the patient assignment between the treatment and control in cluster $k$. This is perhaps more clear in \eqref{eq:pi2ks}. Due to the $r:1$ randomization, the term $(r\pi_{1,k} - \pi_{2,k})$ in \eqref{eq:pi2ks} reflects the difference in the expected sample sizes between treatment and control for cluster $k$. When the term has a larger value, there is a larger difference (imbalance) of information between the two arms, which leads to a large $\pi_{2,k}^*$, and therefore larger $\alpha_k$. In other words, when the treatment arm has more patients than the control arm, PAM-HC borrows more to augment the control. 
\subsection{Estimate treatment effects}
Conditional on $\bm{Z}$, the cluster membership, we assume treatment effects are cluster specific. 
Due to randomization, we assume $C_{1,2} = O_1 = O_2$, i.e., the treatment and control arms share all the clusters  and there are no unique clusters in each of the two arms. 
Under this setting, denote $\Delta_k$ the cluster-specific treatment effect, for $k \in C_{1,2}$, given by
$$\Delta_k = \theta_{1,k} - \theta_{2,k}.$$
In rare cases where there are unique clusters in the treatment or control arms,  we use an ad-hoc rule to merge the unique clusters to a common cluster in $C_{1,2}$ that has the shortest distance in terms of L2-norm between the cluster means.
The overall treatment effect can be computed as a weighted average of the cluster-specific treatment effects $\Delta_k$. 
Conditional on $\bm{Z}$,  we let
\begin{equation} \label{eq:OE}
    \Delta(\bm{Z}) = \sum_{k \in O_1} \pi_{1,k} \Delta_k = \sum_{k \in O_1} \pi_{1,k}(\theta_{1,k} - \theta_{2,k})
\end{equation}
be the  conditional overall treatment effects. We could either use $\{\pi_{1,k}\}$ or $\{\pi_{2,k}\}$ as the weights, which are in principle close to each other due to randomization. However, we decide to use $\{\pi_{1,k}\}$ since the treatment arm (group $j=1$) is expected to have more patients and therefore lead to more stable estimates of clustering weights. 
The (unconditional) overall treatment effect is given by $\Delta = E[\Delta(\bm{Z})].$

\subsection{Inference} \label{subsec:inf}

\cite{bi2023pam} develop a slice sampler to generate posterior samples via Markov chain Monte Carlo (MCMC) simulations. We use $m=1, \ldots, M$ to index the $M$ MCMC samples and use a generic notation $\hat{X}^{(m)}$ to denote the $m$-th sample for random variable $X$. Also, for simplicity, 
let $\bm{D} = (\bm{y}_1,\bm{x}_1, \bm{y}_2,\bm{x}_2,\bm{y}_3,\bm{x}_3)$ denote the entire data, including the data from the RCT and external source. 
Note that the posterior mean of overall treatment effect can be expressed as an integration of \eqref{eq:OE} over the posterior distributions of  $\theta$'s, $\pi$, and $\bm{Z}$, i.e., 
\begin{dmath} \label{eq:post_te}
    \text{E}[\Delta| \bm{D} ] \equiv \text{E}[\text{E}[\Delta(\bm{Z})|\bm{Z},\bm{D}]] \\ = \int \left\{ \int \sum_{k \in O_1}\pi_{1,k}(\theta_{1,k} - \theta_{2,k})p(\theta_{1,k}|\bm{D}, \bm{Z}) p(\theta_{2,k}|\bm{D}, \bm{Z}) p(\bm{\pi}_1|\bm{D}) d\bm{\theta} d\bm{\pi}_1  \right\} p(\bm{Z} |  \bm{D}) d\bm{Z}
\end{dmath} 
where $O_1$, $\pi_{1,k}$ and $k$ are all functions of $\bm{Z}$, $\bm{\theta}=\{(\theta_{1,k}, \theta_{2,k}): k \in O_1\}$, and $\bm{\pi}_1=\{\pi_{1,k}: k \in O_1\}$. 
Using the MCMC samples,  the posterior mean \eqref{eq:post_te} is estimated as follows. For the $m$-th sample, let $\hat{O}_j^{(m)}$ be the  cluster labels for group $j$
. Let cluster $k^{(m)} \in \hat{O}_1^{(m)}$, then 
we compute the $m$-th  posterior sample of the cluster-specific treatment effect 
as
$$\hat{\Delta}_{k^{(m)}}^{(m)} = \hat{\theta}_{1,{k^{(m)}}}^{(m)} - \hat{\theta}_{2,{k^{(m)}}}^{(m)}.$$
Finally, 
the overall treatment effect $\hat{\Delta}^{(m)}$ can be obtained with $\hat{\Delta}_{k^{(m)}}^{(m)}$ and the weights $\hat{\bm{\pi}}_1^{(m)}$ using equation \eqref{eq:OE}:
$$\hat{\Delta}^{(m)} = \sum_{k^{(m)} \in \hat{O}_1^{(m)}} \hat{\pi}_{1,k^{(m)}}^{(m)}\hat{\Delta}_{k^{{(m)}}}^{(m)},$$
and the posterior mean treatment effect is estimated as $\sum_{m=1}^M \hat{\Delta}^{(m)}/M.$ Also, given the posterior sample $\{\hat{\Delta}^{(m)}, \; m=1, \ldots, M\}$,  we can easily compute various quantities of interest, such as the standard deviation of the overall treatment effect. Additionally, it allows us to determine the posterior probability of a significant treatment effect, denoted as 
$$\text{Pr}(\Delta > \epsilon|\text{Data}),$$
for some minimal clinically meaningful treatment effect  $\epsilon$. 

\section{Simulation Studies}
\label{sec:simulation}

\subsection{Simulation Setup}

We generate covariates values $\bm{x}_{i,1}$ and $\bm{x}_{i,2}$ for the current RCT by simulating from a mixture of three multivariate normal distributions. Specifically, 
$$\bm{x}_{i,j} \sim 0.3\times MVN(\bm{2}_3, \bm{I}) + 0.4\times MVN(\bm{0}_3, \bm{I}) + 0.3\times MVN(\bm{-2}_3, \bm{I}), \,\, j = 1, 2,$$
where notation $\bm{a}_3 = [a,a,a]^T$ and $\bm{I}$ is the 3 by 3 identity matrix. 
We generate the covariates of the external data, denoted as $\bm{x}_{i,3}$, under three different scenarios:
\begin{itemize}
    \item Scenario 1 (Superset): we introduce an \textit{additional cluster} with a cluster mean of $\bm{-4}_3$ and a covariance matrix of $\bm{I}$. The weights assigned to the clusters are also different from those used in the RCT:
    $$\bm{x}_{i,3} \sim 0.2\times MVN(\bm{2}_3, \bm{I}) + 0.3\times MVN(\bm{0}_3, \bm{I}) + 0.3\times MVN(\bm{-2}_3, \bm{I}) + \underbrace{0.2\times MVN(\bm{-4}_3, \bm{I})}_{\text{unique}}.$$
    \item Scenario 2 (Overlap): the external data shares some clusters with the current RCT while also having a unique cluster. Specifically, we \textit{remove} the cluster with 
    mean  $\bm{-2}_3$ 
    from the RCT, and similar to Scenario 1, we \textit{add a cluster} with  mean  $\bm{-4}_3$ and covariance $\bm{I}$ to the external data:
    $$\bm{x}_{i,3} \sim 0.5\times MVN(\bm{2}_3, \bm{I}) + 0.3\times MVN(\bm{0}_3, \bm{I}) + \cancel{0.3\times MVN(\bm{-2}_3, \bm{I})} + \underbrace{0.2\times MVN(\bm{-4}_3, \bm{I})}_{\text{unique}}.$$
    \item Scenario 3 (Subset): In this scenario, the external data is a subset of the current RCT. Similar to Scenario 2,  we \textit{remove} the cluster with mean  $\bm{-2}_3$ 
    from the RCT, but do not add any clusters to the external data:
    $$\bm{x}_{i,3} \sim 0.5\times MVN(\bm{2}_3, \bm{I}) + 0.5\times MVN(\bm{0}_3, \bm{I}) + \cancel{0.3\times MVN(\bm{-2}_3, \bm{I})}.$$
\end{itemize}

\noindent For outcomes $\bm{y}$, we assume that they are associated with the baseline covariates. We consider both continuous and binary outcomes. 

For continuous outcomes, we simulate the outcome $y_{i,j}$ using a linear regression given by:
$$y_{i,j} \sim \beta_{0,j} + \bm{\beta}^T\bm{x}_{i,j} + \epsilon_{i,j}, \,\, \epsilon_{i,j} \sim N(0,1), \,\, j = 1, 2, 3.$$
For binary outcomes, we generate $P(y_{i,j} = 1|\bm{x}_{i,j})$ using a logistic regression given by:
$$\text{logit } P(y_{i,j} = 1|\bm{x}_{i,j}) = \beta_{0,j} +  \bm{\beta}^T\bm{x}_{i,j},$$
and then simulate $y_{i,j}$ from the Bernoulli distribution with the success probability $P(y_{i,j} = 1|\bm{x}_{i,j})$. 
We fix $\beta_{0,2} = \beta_{0,3} = 0$, and set $\bm{\beta} = \bm{1}_3$. We consider four cases $\beta_{0,1} = 0, \, 1, \, 2, \, \text{or } 3$, which is the intercept (also treatment effect) in the outcome regression models in the treatment arm. 
Furthermore, we assume that the current RCT uses a randomization ratio of $r=2:1.$ We simulate the RCT data with  two different sample sizes, $N = \{300, 450\}$. For the case where $N = 300$, we place  200 patients in the treatment arm and 100  in the control.  We generate a total of 300 patients for the external data. For the case where $N = 450$, we place 300, 150, and 450 patients in the treatment arm, control arm, and the external data, respectively.

The parameter of interest is the treatment effect $\Delta$, which represents the difference in mean responses between the treatment and control arms. Consequently to $\beta_{0,1} = 0, \, 1,\, 2, \, \text{and } 3,$ the true values of $\Delta$ are 0, 1, 2, and 3, respectively, for the continuous outcome, and 
$0.00\%$, $9.42\%$, $15.76\%$, and $19.52\%$ 
, respectively, for the binary outcome. 



We compare the proposed PAM-HC method with two other methods. The first method, referred to as the baseline method, estimates the treatment effect using only the RCT data and does not borrow information from the external data. Inference of treatment effect is based on the maximum likelihood estimation (MLE). 
The second method is the PSCL method proposed by \cite{chen2020propensity}. We implement the PSCL method using three different strategies, corresponding to the ways propensity scores are estimated. Namely, they are PSCL 1, PSCL 2, and PSCL 3, referring to the PSCL method with propensity scores estimated using a first-order logistic regression, random forest, and a logistic regression with model selection, respectively. All three versions of PSCL are implemented in the R package ``PSRWE" \citep{wang2022psrwe}. Finally, for  each scenario, we simulate 100 datasets. 

We use the following priors and hyperparameters for the proposed PAM-HC method. For PAM model (\ref{eq:PAM}), we set $a = b = 0.5$, and utilize a $Gamma(3,3)$ prior for the hyperparameters $\gamma$ and $\alpha_0$. Additionally, we set $\bm{\mu}_0 = \bm{0}_3$, $\Psi = \bm{I}$, $\lambda = 0.1$, and $\nu = 3$. For the power priors in PAM-HC, we adopt a normal distribution $N(\mu,\sigma^2)$ for the continuous outcome and a Bernoulli distribution $\text{Bern}(p)$ for the binary outcome. Conjugate priors are chosen for the parameters in the sampling model. Specifically, we use a normal-inverse-gamma prior $NIG(0, 0.1, 3, 3)$ for $(\mu,\sigma^2)$ and $Beta(0.5, 0.5)$ for $p$. These are standard priors that are not informative and routinely applied in the literature. Lastly, we run an MCMC simulation of 10,000 iterations, with burn-in period of 5,000 iterations. 

We summarize the posterior cluster membership using an optimal clustering method \citep{meilua2007comparing} to obtain a point estimate. 
To assess the clustering accuracy in comparison to the ground truth cluster membership of each patient, we use the adjusted Rand index (ARI) \citep{hubert1985comparing} and the normalized Frobenius distance (NFD) \citep{horn1990norms}. More detail can be found in \cite{bi2023pam}. Lastly, we 
assess  the performance of PAM-HC  in terms of  the estimated overall treatment effect, including the mean, standard deviation, bias, and mean squared error (MSE) across all simulated datasets. 


\subsection{Simulation results}

\begin{figure}[!p]
\begin{center}
\includegraphics[width=0.8\textwidth]{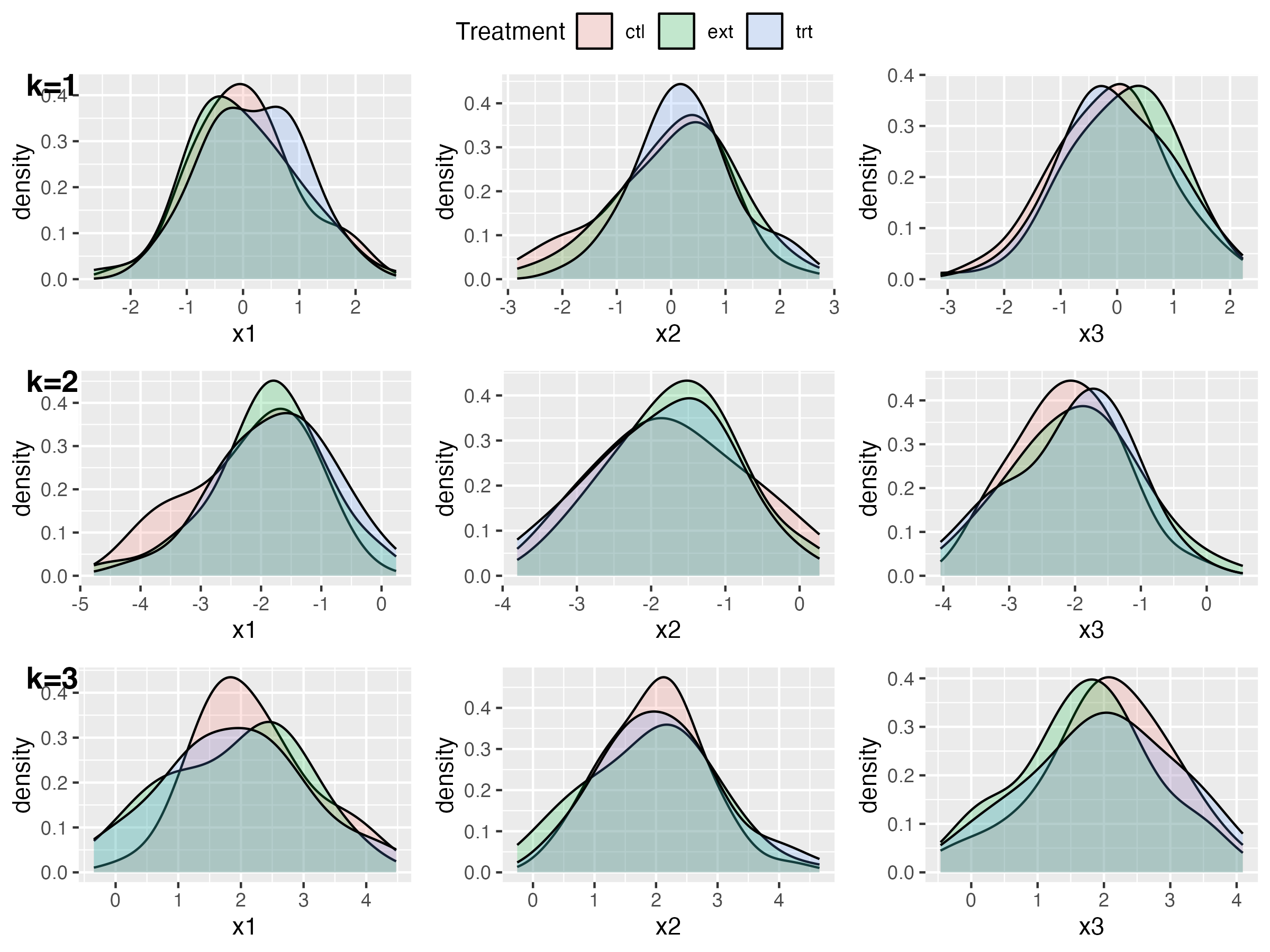}
\end{center}
\caption{The covariate density plots of one simulated data in Scenario 1. The rows represent three clusters estimated by PAM-HC.}
\label{fig:cov_balance_plot_sc1}
\end{figure}

We assess similarity in the distributions of covariates between the treatment and HC.
We first check that under PAM-HC, if the distribution of covariates of the treatment arm is similar to the distribution of the hybrid control arm. Specifically, within each estimated cluster, the distributions of covariates should be similar between the two arms. 
We randomly selected one dataset in each scenario with $N = 300$. We plot the density of the covariates by estimated clusters in Figure \ref{fig:cov_balance_plot_sc1} below for Scenario 1, and in Figures \ref{fig:cov_balance_plot_sc2} and \ref{fig:cov_balance_plot_sc3} in Appendix for scenarios two and three, respectively. 

Figure \ref{fig:cov_balance_plot_sc1} shows that the distribution of each covaraite are indeed similar between the treatment arm, control arm, and the external data, for each inferred cluster $k$. Furthermore, the estimated cluster centers are shown in Table \ref{tab:clusters_center} below. In addition, the cluster-specific treatment effects for the three selected datasets 
are reported in Tables \ref{tab:cluster_cont} and \ref{tab:cluster_bin} in Appendix. In the selected examples, we see that PAM-HC is able to correctly identify the number of clusters in these selected examples. The estimated cluster centers are also close to the true values in their corresponding scenarios. In addition, Tables \ref{tab:cluster_cont} and \ref{tab:cluster_bin} suggest that treatment effects for clusters are well estimated. 

To further assess the similarity of covariate distributions between treatment and the hybrid control arms, we follow the procedure outlined in \cite{chandra2021bayesian} and apply the Bayesian Additive Regression Tree (BART) model \citep{chipman2010bart}. 
For each estimated 
cluster $k$, we aggregate data from all three groups and create  a dummy variable $T_i$ indicating whether patient $i$ belongs to the treatment arm ($T_i = 1$) or the hybrid control arm ($T_i = 0$). We then carry out a 10-fold cross-validation, with 9-folds used as the training data and 1-fold as the testing data. We apply BART to predict whether an observation in the testing data belongs to the treatment arm or not. The results are reported as the Area Under the ROC Curve (AUC). A value around 0.5 indicates no difference between the patients in the current treatment arm and the patients in the hybrid control. We randomly select 10 datasets in each scenario, and the corresponding results 
show that across all scenarios, the range of mean AUC values is between 0.525 and 0.544, all 
around 0.5. This indicates that the covariate distributions for the treatment and hybrid control arms are similar and indistinguishable by BART. 


Next, we report the clustering results of PAM-HC for all simulated datasets. The true number of clusters is four in Scenario 1 and Scenario 2,  and three in Scenario 3. 
For ARI and NFD, the closer the value of ARI is to 1 or the value of NFD  to 0, the better the clustering result of the method. Table \ref{tab:cluster} in Appendix shows the estimated total number of clusters across all groups, as well as the ARI and the NFD of the estimated clusters compared to the true cluster membership.
On average, the number of estimated cluster is accurate, close to its truth in all cases. 
The ARI and NFD values are  satisfactory, improving with increasing sample size. 

Lastly, we present the estimated treatment effect, 
its standard deviation, 
and the mean squared error (MSE). We compare these results with the baseline method and PSCL 1-3. 
Table \ref{tab:continuous_result} in Appendix provides a summary of the results. 
In the case of a smaller sample size ($N = 300$), PAM-HC shows the lowest MSE in Scenario 2 and lowest bias in Scenario 3, for all four values of $\Delta$. The performance of the PAM-HC design improves further with a larger sample size ($N = 450$). 
PAM-HC is also comparable to the PSCL methods in MSE and much smaller than the baseline method in Scenarios 1 and 2. We also evaluate the performance of PAM-HC with binary outcomes, 
and the results are shown in Table \ref{tab:bin_result} in Appendix. Similar to the continuous outcome, PAM-HC exhibits desirable performance. 

Scenario 2 is an interesting case in which some but not all clusters are shared across groups. This is where PAM-HC excels in its performance.
Since PAM-HC is designed to capture the pattern of overlapping clusters, it leads to
more precise information borrowing and better performance.
To see this, we assess the ``inclusion probability", defined as the probability of each patients in the external data being borrowed for HC. Mathematically, this is equal to $\mbox{Pr}(Z_{i,3} \notin C_{2,3} \mid  \bm{D})$. In words, if  a patient $i$ in the external data group ($j = 3$) is not in a common cluster with the control, the patient is not  ``borrowed" for forming the HC. 
In Scenario 2, for patients in the unique cluster 4 in external data, \color{black} the mean and SD inclusion probability across the patients are 3.47\% (SD = 0.17) and 2.51\% (SD = 0.15) for sample sizes $N =$ 300 and 450, respectively. For patients in other common clusters in the external data, the inclusion probability are all greater than 91\%. These results demonstrate that PAM-HC is able to adaptively borrow based on the overlapping status of each cluster.

\section{Application}
\label{sec:application}
\subsection{Background and Dataset}

We consider clinical trials for patients with Atopic Dermatitis (AD). 
AD is a significant contributor to skin-related disability globally, characterized by recurrent eczematous lesions and intense itch \citep{simpson2022efficacy}. 
In this application, 
we analyze data from the control arms (placebo) of three historical trials (with NCT numbers NCT03569293, NCT03607422 and NCT03568318) for AD. The treatment arms and their data are not available for analysis due to confidentiality.  
Specifically, the three control arms of the three historical trials  share similar inclusion and exclusion criteria, and the patients in the three control arms all receive a placebo. The control arms consist of 263, 265, and 306 patients. 

Each trial reports several baseline characteristics of the patients, including their gender, age, race, ethnicity, body mass index (BMI), baseline body surface area affected (BSA), and baseline Eczema Area and Severity Index (EASI). Additionally, the trials record the EASI score at the 16-week mark to assess the progression of the patients' disease. The primary outcome is the percent change in the EASI score from baseline to 16 weeks, denoted as:
$$\text{EASI}_{\text{pc}} = \frac{\text{Baseline EASI} - \text{16-week EASI}}{\text{Baseline EASI}} \times 100\%$$
The binary response to the treatment is defined as $\text{EASI}_{\text{pc}} \geq 75\%$. In other words, the binary outcome, denoted as $y$, is defined as
$$y = \left\{ \begin{matrix} 1, & \text{if } \text{EASI}_{\text{pc}} \geq 75\%; \\ 0 & \text{otherwise.} \end{matrix} \right .$$
To illustrate PAM-HC, we pretend the control arm of trial one is the treatment arm of a hypothetical RCT and randomly select 131 patients from the control arm of trial two to serve as the RCT control. Therefore, we construct a hypothetical RCT of 394 patients 
with a randomization ratio of $2:1$. We examine the distributions of the covaraites between the hypothetical treatment and control arms and find no major differences (results not shown).  Lastly, we use 
the 306 patients of trial three as the external data with which we build a hybrid control for the RCT. The observed 
response rates 
are of 25\%, 21\%, and 34\%, for trials one, two, and three, respectively. And the overall mean responses is roughly 28\% across all three trials. We use these data for PAM-HC in a null scenario. 

Alternatively, to construct a trial with an actual treatment effect (the alternative case), we follow the findings of \cite{simpson2022efficacy} that reports a response rate of 80\% under the treatment arm. We spike in response data in trial one and use it as the treatment arm in the hypothetical RCT. Specifically, we generate a treatment arm (consisting of the 263 patients from trial one) based on the following procedure. To make sure that the outcome is related to the covariates, we first fit a logistic regression model with the original outcome of trial one ($y_{i,1}$) as the dependent variable and the four covariates ($\bm{x}_{i,1}$) as the independent variables. We then fixed the estimated regression coefficients $\hat{\bm{\beta}}$ and conducted a grid search to find a value of $\tilde{\beta}_{0,1} = 2.33$ that satisfied the condition 
$$\tilde{y}_{i,1} \sim Bern(p_i), \,\,  \text{logit } p_i = \hat{\bm{\beta}}^T\bm{x}_{i,1} + \tilde{\beta}_{0,1},$$ 
$\tilde{\bm{y}}_1 = \{\tilde{y}_{i,1}\}_{i=1}^{263}$, and $\text{Pr}(\tilde{\bm{y}}_1 = 1) \approx 80\%$. 
The true treatment effect is roughly $80\% - 28\% \approx 52\%$ after spike-in. These data form the alternative scenario. 



\subsection{Analysis Results}
The posterior mean number of clusters by PAM-HC is 4.15 (SD = 0.36), and
PAM-HC generates a point-estimate of cluster structure that consists of  four common clusters that are shared across all three arms without a unique cluster. 
Table \ref{tab:application_cluster_specific} below summarizes the cluster means as well as the cluster-specific treatment effects of PAM-HC for the null and alternative scenarios. 

We report the estimated treatment effects using  the PAM-HC method as well as the baseline and PSCL 1-3 methods. The results are shown in Table \ref{tab:app_result}. We observe that all methods report small treatment effects that are not statistically significant under the null case. However, when borrowing information from external data, the PAM-HC and PSCL methods report negative treatment effects as opposed to a positive treatment effect reported by the baseline method which does not borrow information from external data. This is expected since the response rate of the external data is 34\%, higher than those of the hypothetical treatment (25\%) and control (21\%) arms.  For the alternative case, all methods find non-zero treatment effects, although PAM-HC and the PSCL methods report a lower treatment effect compared to the baseline. Again, this is expected since when borrowing from the external data with  34\% response, the control response rate is expected to increase from 21\% in the hybrid control. 
Lastly, the proposed PAM-HC method reports an accurate estimation of the treatment effect in the alternative case, which is around the ground true of 52\%. 

\section{Discussion}
\label{sec:conclusion}

In this study, we introduce the PAM-HC method to augment the control arm of an RCT using external data and improve the estimation of treatment effects. A key innovation is to identify  
common subpopulations of patients between the  RCT and the external data and allow information to be borrowed only across these common subpopulations. 
We find that PAM-HC  performs well when compared to existing methods in the simulation and case study, especially when not all patient subpopulations are shared between RCT and external data. Thanks to the model-based inference on all the unknown parameters using BNP models, PAM-HC is powerful in reporting posterior distributions of cluster-specific treatment effects, overall treatment effects, and the random clusters themselves. 

However, it is important to acknowledge the limitations of the current method. Firstly, the assumption that covariates are continuous variables restricts the applicability of PAM to handle binary and categorical variables. 
Another limitation lies in the underlying assumption that the covariates used for clustering and inference includes all the relevant confounders. Future work is ongoing to address these issues.

\vskip 2in 

\section*{Disclaimer}

The external control data was based in part on data from the TransCelerate BioPharma Inc.
Historical Trial Data (HTD) Sharing Initiative, which includes contributions of anonymized or
pseudonymized data from TransCelerate HTD member companies including AbbVie, Amgen,
Astellas, AstraZeneca, Boehringer Ingelheim, Bristol-Myers Squibb, Eli Lilly, GlaxoSmithKline,
Johnson \& Johnson, Merck KGaA, Novartis, Novo Nordisk, Pfizer, Roche, Sanofi,
Shionogi, and UCB Pharma (``Data Providers").
Neither TransCelerate Biopharma Inc. nor the Data Providers have contributed to or approved or are in any way responsible for this research result.

\bibliographystyle{plainnat}
\bibliography{reference}

\begin{table}[!p]
    \caption{Cluster mean estimated by PAM-HC on selected examples from each of the three scenarios. The entries for columns $x_1$, $x_2$, and $x_3$ are posterior $\text{mean}_{\text{SD}}$ (truth), and estimated clusters (truth) for the last column.}
    \label{tab:clusters_center}
    \centering
    \begin{tabular}{c|ccccc}
    \hline
       \multirow{2}{3em}{Sc} & \multirow{2}{3em}{Cluster}  & \multicolumn{3}{c}{Cluster mean} & \multirow{2}{6em}{Groups ($j$)} \\
       & & $x_1$ & $x_2$ & $x_3$ & \\
    \hline
       \multirow{4}{3em}{Sc 1}  & 1 & $0.08_{0.91}$ (0) & $0.14_{1.03}$ (0) & $0.01_{0.96}$ (0) & 1,2,3 (1,2,3)  \\
       & 2 & $-1.89_{0.95}$ (-2) & $-1.72_{0.88}$ (-2) & $-2.03_{0.90}$ (-2) & 1,2,3 (1,2,3)  \\
       & 3 & $1.98_{1.05}$ (2) & $1.99_{0.93}$ (2) & $1.91_{1.02}$ (2) & 1,2,3 (1,2,3) \\
       & 4 & $-4.13_{0.96}$ (-4) & $-4.03_{0.90}$ (-4) & $-4.15_{0.96}$ (-4) & 3 (3) \\
    \hline
       \multirow{4}{3em}{Sc 2}  & 1 & $0.01_{0.90}$ (0) & $0.19_{1.07}$ (0) & $-0.01_{0.95}$ (0) & 1,2,3 (1,2,3) \\
       & 2 & $-1.88_{0.97}$ (-2) & $-1.95_{1.14}$ (-2) & $-1.96_{0.81}$ (-2) & 1,2 (1,2) \\
       & 3 & $2.02_{0.98}$ (2) & $2.05_{1.11}$ (2) & $1.94_{0.98}$ (2) & 1,2,3 (1,2,3) \\
       & 4 & $-3.85_{1.05}$ (-4) & $-4.08_{0.90}$ (-4) & $-3.88_{0.85}$ (-4) & 3 (3) \\
    \hline
       \multirow{3}{3em}{Sc 3}  & 1 & $-0.09_{0.92}$ (0) & $-0.66_{0.75}$ (0) & $-0.38_{0.93}$ (0) & 1,2,3 (1,2,3)  \\
       & 2 & $-1.95_{0.90}$ (-2) & $-2.00_{1.12}$ (-2) & $-1.99_{0.81}$ (-2) & 1,2 (1,2) \\
       & 3 & $1.29_{1.38}$ (2) & $1.61_{1.10}$ (2) & $1.33_{1.28}$ (2) & 1,2,3 (1,2,3) \\
    \hline
    \end{tabular}
\end{table}

\begin{table}[!p]
\caption{Estimated cluster mean and cluster-specific treatment effect using the data of the AD trial. The entries are posterior $\text{mean}_{\text{SD}}$.}
\label{tab:application_cluster_specific}
    \centering
    \resizebox{\linewidth}{!}{
    \begin{tabular}{c|ccc|cccc|cc}
    \hline
       \multirow{2}{5em}{Cluster $k$} & \multicolumn{3}{|c}{Weights} & \multicolumn{4}{c}{Cluster mean} & \multicolumn{2}{c}{Cluster-specific treatment effect} \\
        & $\pi_{1,k}$ & $\pi_{2,k}$ & $\pi_{3,k}$ & Age & Baseline EASI & BSA & BMI & Null & Alternative \\ 
    \hline
       Cluster 1 & 0.14 & 0.18 & 0.16 & $19.86_{0.70}$  & $18.55_{0.46}$ & $24.56_{1.07}$ & $22.78_{0.50}$ & $0.077_{0.113}$ & $0.659_{0.102}$ \\
       Cluster 2 & 0.20 & 0.18 & 0.23 & $37.79_{1.37}$  & $39.04_{1.53}$ & $66.30_{2.52}$ & $29.11_{0.79}$ & $-0.031_{0.074}$ & $0.561_{0.076}$ \\
       Cluster 3 & 0.42 & 0.40 & 0.32 & $41.26_{1.54}$  & $21.68_{0.40}$ & $31.79_{1.26}$ & $27.30_{0.74}$ & $-0.081_{0.068}$ & $0.500_{0.065}$ \\
       Cluster 4 & 0.24 & 0.24 & 0.29 & $21.56_{0.88}$  & $33.82_{1.87}$ & $57.06_{2.85}$ & $22.12_{0.40}$ & $-0.080_{0.091}$ & $0.519_{0.091}$ \\
    \hline
    \end{tabular}
    }
\end{table}

\begin{table}[!p]
\caption{Estimated treatment effects for the AD trial, using the proposed PAM-HC method, the baseline method, and three versions of PSCL method.}
\label{tab:app_result}
\begin{center}
\begin{tabular}{ cc|cc }
    \hline
    & Method & $\Delta_{\text{SD}(\Delta)}$ & Significance \\
    \hline
    \multirow{5}{6em}{Null Case} &Baseline & $0.026_{0.044}$ & P-value: 0.281 \\
    & PSCL1 & $-0.045_{0.149}$ & P-value: 0.618 \\
    & PSCL2 & $-0.034_{0.149}$ & P-value: 0.590\\
    & PSCL3 & $-0.041_{0.315}$ & P-value: 0.551 \\
    & PAM-HC & $-0.049_{0.037}$ & $\text{Pr}(\Delta > 0|\text{data}) = 0.09$ \\
    \hline
    \multirow{5}{6em}{Alternative Case} & Baseline & $0.627_{0.042}$ & P-value: 0.000 \\
    & PSCL1 & $0.596_{0.149}$ & P-value: 8.4e-6 \\
    & PSCL2 & $0.606_{0.139}$ & P-value: 6.5e-6 \\
    & PSCL3 & $0.598_{0.140}$ & P-value: 1.0e-5 \\
    & PAM-HC & $0.539_{0.036}$ & $\text{Pr}(\Delta > 0|\text{data}) = 1.00$ \\
    \hline
\end{tabular}
\end{center}
\end{table}

\clearpage
\newpage
\appendix

\section{}

\renewcommand\thefigure{\thesection.\arabic{figure}} 
\setcounter{figure}{0}

\renewcommand\thetable{\thesection.\arabic{table}} 
\setcounter{table}{0}

\renewcommand\theequation{\thesection.\arabic{equation}} 
\setcounter{equation}{0}

\subsection{Additional Simulation Results}

\begin{figure}[!p]
\begin{center}
\includegraphics[width=0.8\textwidth]{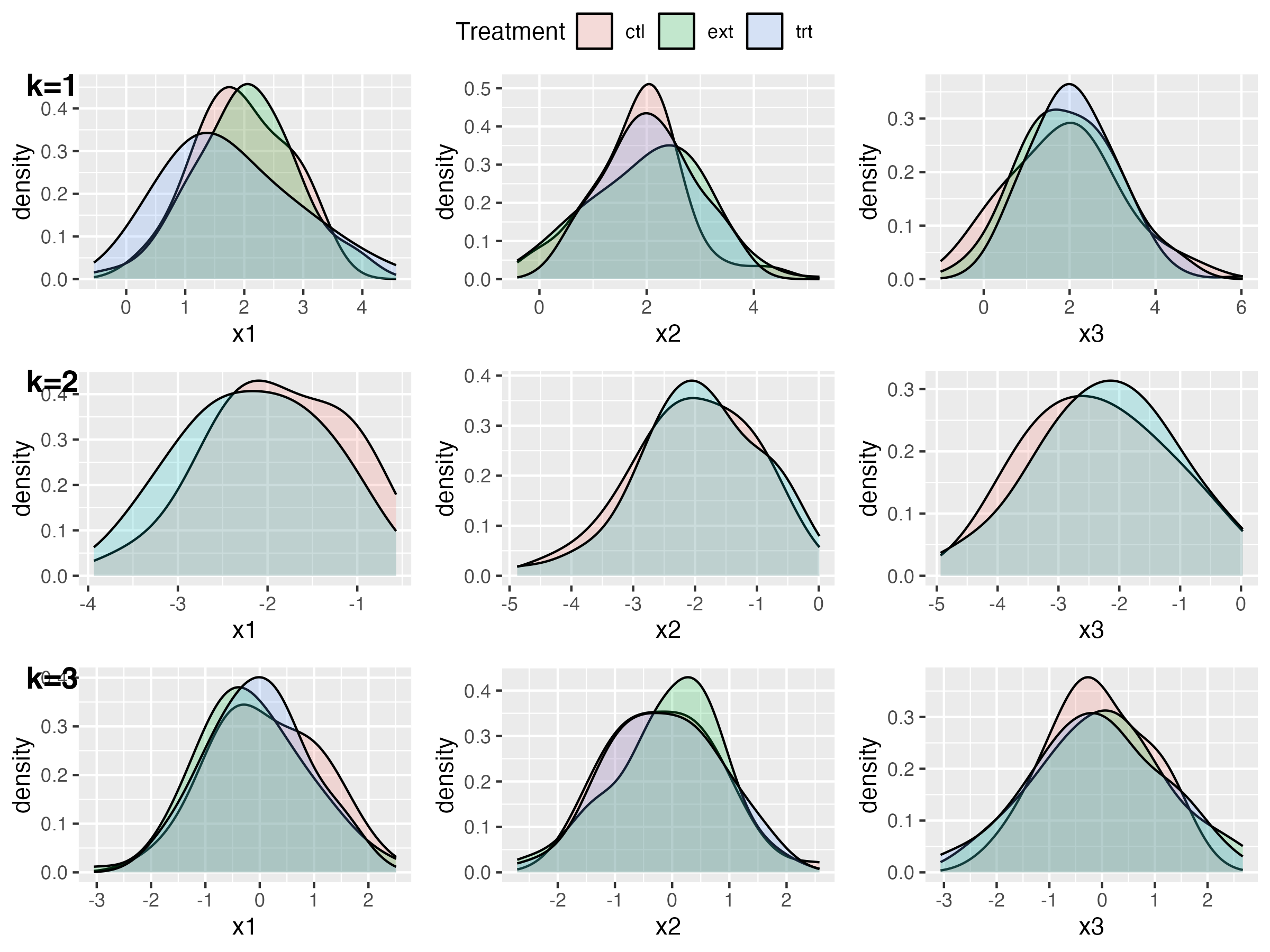}
\end{center}
\caption{The covariate density plots of one simulated data in Scenario 2. The rows represent three clusters estimated by PAM-HC.}
\label{fig:cov_balance_plot_sc2}
\end{figure}

\begin{figure}[!p]
\begin{center}
\includegraphics[width=0.8\textwidth]{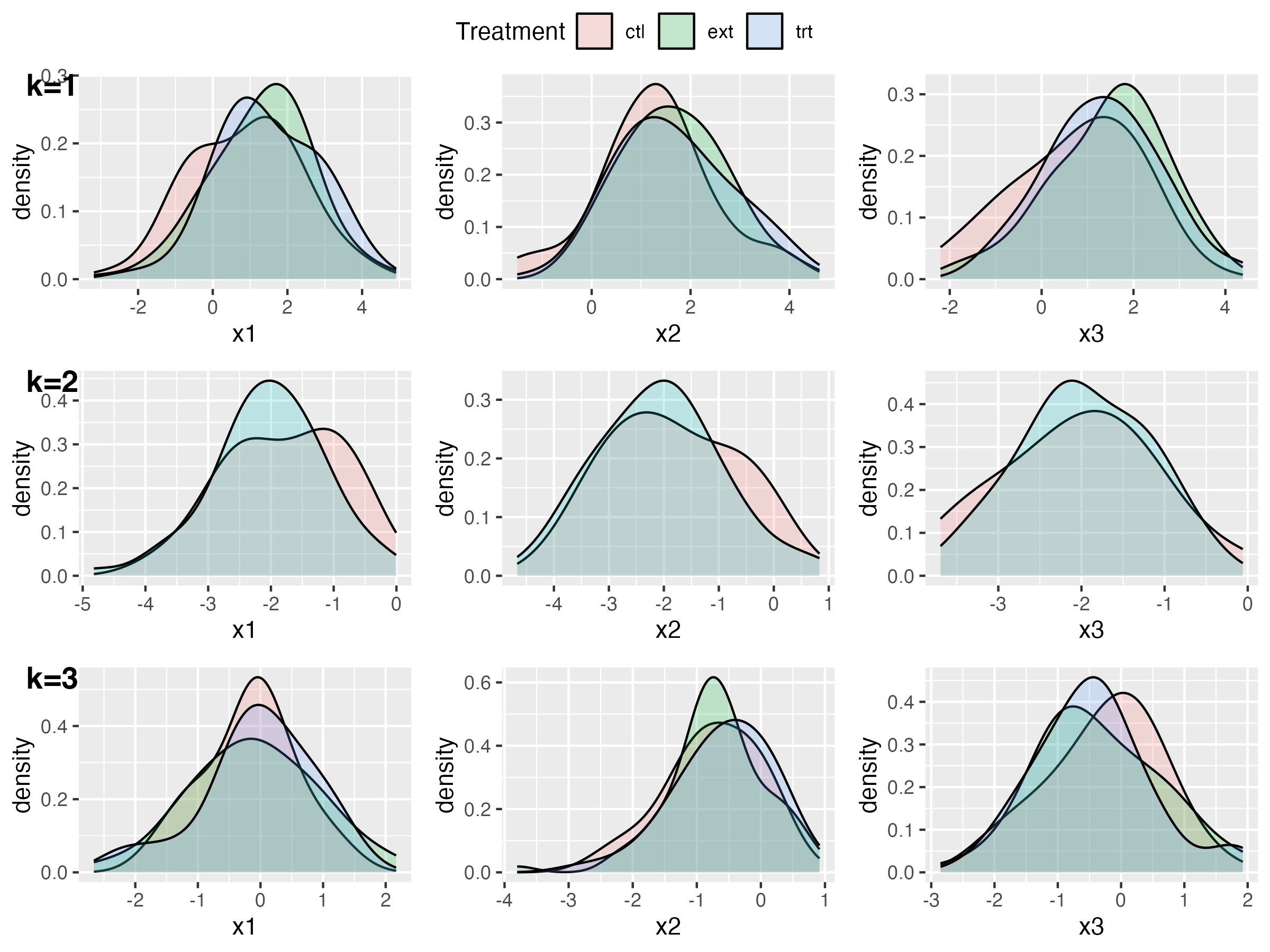}
\end{center}
\caption{The covariate density plots of one simulated data in Scenario 3. The rows represent three clusters estimated by PAM-HC.}
\label{fig:cov_balance_plot_sc3}
\end{figure}

Figures 
\ref{fig:cov_balance_plot_sc2} and \ref{fig:cov_balance_plot_sc3} are the density plots of covariates for the randomly selected example dataset from Scenarios 2 and 3, respectively. Each row represents a cluster, each column represents a dimension of the multivariate covariate, and each color represent a different treatment group (treatment arm, control arm, or the external data).


Tables \ref{tab:cluster_cont} and \ref{tab:cluster_bin} show the cluster-specific treatment effects for each scenario for the continuous and binary outcomes, respectively.

\begin{table}[!p]
\caption{Estimated cluster-specific treatment effects for selected examples with different values of true $\Delta$ in each of the three scenarios using continuous outcome in the simulation. The entries for the three columns, Cluster 1, Cluster 2, and Cluster 3, are posterior $\text{mean}_{\text{SD}}$ (observed cluster-specific treatment effects). The entries for the last column are posterior $\text{mean}_{\text{SD}}$ (truth).}
    \label{tab:cluster_cont}
    \centering
    \resizebox{\linewidth}{!}{
    \begin{tabular}{c|ccc|c}
    \hline
       \multirow{2}{3em}{Sc} & \multicolumn{3}{|c|}{Cluster specific treatment effect} & \multirow{2}{7em}{Overall treatment effect} \\
       & Cluster 1 & Cluster 2 & Cluster 3 & \\
    \hline
       \multirow{4}{3em}{Sc 1} & $0.18_{0.18}$ (0.03) & $-0.09_{0.21}$ (0.01) & $0.23_{0.18}$ (0.26) & $0.12_{0.11}$ (0) \\
       & $1.27_{0.18}$ (1.08) & $-0.02_{0.22}$ (0.05) & $2.22_{0.18}$ (2.24) & $1.22_{0.12}$ (1) \\
       & $1.99_{0.18}$ (1.97) & $0.81_{0.21}$ (0.86) & $2.62_{0.19}$ (2.60) & $1.87_{0.12}$ (2) \\
       & $3.14_{0.17}$ (3.14) & $2.04_{0.21}$ (2.02) & $4.17_{0.19}$ (4.13) & $3.17_{0.12}$ (3) \\
    \hline
       \multirow{4}{3em}{Sc 2}  & $-0.10_{0.17}$ (-0.01) & $-0.16_{0.23}$ (-0.19) & $-0.14_{0.23}$ (-0.14) & $-0.13_{0.11}$ (0) \\
       & $0.95_{0.18}$ (0.93) & $-0.08_{0.24}$ (-0.22) & $1.87_{0.22}$ (1.82) & $0.76_{0.13}$ (1) \\
       & $1.99_{0.20}$ (1.83) & $1.33_{0.21}$ (1.28) & $3.16_{0.21}$ (3.10) & $2.01_{0.13}$ (2) \\
       & $3.30_{0.19}$ (3.24) & $2.04_{0.21}$ (1.91) & $4.04_{0.22}$ (3.94) & $2.98_{0.12}$ (3) \\
    \hline
       \multirow{4}{3em}{Sc 3}  & $-0.11_{0.23}$ (-0.12) & $-0.18_{0.23}$ (-0.19) & $-0.11_{0.17}$ (0.03) & $-0.13_{0.11}$ (0) \\
       & $0.74_{0.24}$ (0.82) & $-0.20_{0.23}$ (-0.22) & $1.52_{0.20}$ (1.99) & $0.76_{0.13}$ (1) \\
       & $1.76_{0.26}$ (1.92) & $1.32_{0.22}$ (1.28) & $2.71_{0.21}$ (3.06) & $2.02_{0.12}$ (2) \\
       & $2.92_{0.25}$ (3.12) & $1.95_{0.21}$ (1.91) & $3.85_{0.20}$ (4.08) & $2.97_{0.13}$ (3) \\
    \hline
    \end{tabular}
    }
\end{table}

\begin{table}[!p]
\caption{Estimated cluster-specific treatment effects for selected examples with different values of true $\Delta$ in each of the three scenarios using binary outcome in the simulation. The entries for the three columns, Cluster 1, Cluster 2, and Cluster 3, are posterior $\text{mean}_{\text{SD}}$ (observed cluster-specific treatment effects). The entries for the last column are posterior $\text{mean}_{\text{SD}}$ (truth).}
    \label{tab:cluster_bin}
    \centering
    \resizebox{\linewidth}{!}{
    \begin{tabular}{c|ccc|c}
    \hline
       \multirow{2}{3em}{Sc} & \multicolumn{3}{|c|}{Cluster specific treatment effect} & \multirow{2}{7em}{Overall treatment effect} \\
       & Cluster 1 & Cluster 2 & Cluster 3 & \\
    \hline
       \multirow{4}{3em}{Sc 1}  & $0.06_{0.09}$ (0.07) & $-0.01_{0.04}$ (0.00) & $0.00_{0.03}$ (0.00) & $0.02_{0.04}$ (0.00) \\
       & $0.25_{0.09}$ (0.23) & $0.04_{0.04}$ (0.03) & $0.01_{0.03}$ (0.01) & $0.11_{0.04}$ (0.09) \\
       & $0.33_{0.08}$ (0.35) & $0.04_{0.05}$ (0.05) & $0.01_{0.02}$ (0.00) & $0.14_{0.03}$ (0.16) \\
       & $0.48_{0.07}$ (0.50) & $0.18_{0.07}$ (0.12) & $0.02_{0.03}$ (0.01) & $0.24_{0.04}$ (0.20) \\
    \hline
       \multirow{4}{3em}{Sc 2}  & $0.04_{0.09}$ (0.01) & $-0.01_{0.03}$ (0.00) & $0.02_{0.03}$ (0.01) & $0.02_{0.03}$ (0.00) \\
       & $0.22_{0.09}$ (0.19) & $0.00_{0.06}$ (0.01) & $0.01_{0.02}$ (0.02) & $0.08_{0.04}$ (0.09) \\
       & $0.36_{0.08}$ (0.33) & $0.09_{0.05}$ (0.08) & $0.02_{0.03}$ (0.01) & $0.16_{0.04}$ (0.16) \\
       & $0.41_{0.08}$ (0.39) & $0.14_{0.07}$ (0.12) & $0.01_{0.02}$ (0.02) & $0.19_{0.04}$ (0.20) \\
    \hline
       \multirow{4}{3em}{Sc 3}  & $-0.09_{0.11}$ (-0.10) & $-0.03_{0.04}$ (-0.04) & $0.08_{0.08}$ (0.01) & $0.00_{0.04}$ (0.00) \\
       & $0.15_{0.12}$ (0.10) & $0.00_{0.03}$ (0.00) & $0.12_{0.08}$ (0.03) & $0.08_{0.04}$ (0.09) \\
       & $0.36_{0.11}$ (0.27) & $-0.01_{0.05}$ (-0.01) & $0.16_{0.08}$ (0.01) & $0.14_{0.04}$ (0.16) \\
       & $0.47_{0.10}$ (0.36) & $0.09_{0.05}$ (0.03) & $0.20_{0.07}$ (0.10) & $0.22_{0.04}$ (0.20) \\
    \hline
    \end{tabular}
    }
\end{table}

Table \ref{tab:cluster} shows the estimated number of clusters, the ARI, and the NFD by PAM for each scenario in the simulation study.

\begin{table}[!p]
\caption{Clustering performance for PAM-HC evaluated according to the number of total detected clusters (truth = 4 clusters for Sc 1 and Sc 2; 3 clusters for Sc 3) based on the estimated optimal clustering, the adjusted Rand index (ARI), and the normalized Frobenius distance (NFD). The entries are $\text{mean}_{\text{SD}}$ over 100 datasets.}
\label{tab:cluster}
\begin{center}
    \begin{tabular}{cc|ccc}
    \hline
    $N$ & Sc & Clusters & ARI & NFD \\
    \hline
    \multirow{3}{3em}{300} & Sc 1 & $3.91_{0.46}$ & $0.74_{0.13}$ & $0.09_{0.06}$ \\
     & Sc 2 & $4.01_{0.46}$ & $0.77_{0.16}$ & $0.08_{0.05}$ \\
     & Sc 3 & $3.03_{0.36}$ & $0.75_{0.16}$ & $0.09_{0.07}$ \\
    \hline
    \multirow{3}{3em}{450} & Sc 1 & $4.04_{0.20}$ & $0.81_{0.03}$ & $0.06_{0.01}$ \\
     & Sc 2 & $4.02_{0.20}$ & $0.84_{0.07}$ & $0.05_{0.02}$ \\
     & Sc 3 & $3.09_{0.32}$ & $0.83_{0.05}$ & $0.06_{0.02}$ \\
    \hline
    \end{tabular}
\end{center}
\end{table}

Tables \ref{tab:continuous_result} and \ref{tab:bin_result} show the estimated overall treatment effects with the baseline model, PSCL, and PAM-HC for each simulation scenario for the continuous and binary outcomes, respectively.

\begin{table}[!p]
\caption{Simulation results based on continuous outcome for PAM-HC, baseline method, and three versions of PSCL methods. Here $\hat{\Delta}$ is the average posterior mean of the overall treatment effect across 100 simulated trials.}
\label{tab:continuous_result}
\begin{center}
\resizebox{\linewidth}{!}{
\begin{tabular}{ c|c|c|ccc|ccc|ccc|ccc }
 \hline
 \multirow{2}{3em}{$N$} & \multirow{2}{2em}{Sc} & \multirow{2}{6em}{Method} & \multicolumn{3}{|c|}{True $\Delta = 0$} & \multicolumn{3}{|c|}{True $\Delta = 1$} & 
 \multicolumn{3}{|c|}{True $\Delta = 2$} &
 \multicolumn{3}{|c}{True $\Delta = 3$} \\
  & & & $\hat{\Delta}$ & SD & MSE & $\hat{\Delta}$ & SD & MSE & $\hat{\Delta}$ & SD & MSE & $\hat{\Delta}$ & SD & MSE \\
 \hline
 \multirow{15}{3em}{300} & \multirow{5}{2em}{Sc 1} & Baseline & -0.08 & 0.58 & 0.35 & 0.92 & 0.56 & 0.32 & 1.92 & 0.58 & 0.35 & 2.92 & 0.56 & 0.32 \\
 & & PSCL1 & 0.01 & 0.20 & 0.04 & 1.02 & 0.17 & 0.03 & 2.01 & 0.19 & 0.04 & 3.02 & 0.17 & 0.03 \\
 & & PSCL2 & 0.44 & 0.45 & 0.40 & 1.46 & 0.44 & 0.40 & 2.44 & 0.44 & 0.40 & 3.45 & 0.43 & 0.39 \\
 & & PSCL3 & 0.08  & 0.33 & 0.12 & 1.09 & 0.31 & 0.11 & 2.08 & 0.33 & 0.12 & 3.09 & 0.31 & 0.11  \\
 & & PAM-HC & 0.03 & 0.31 & 0.10 & 1.04 & 0.30 & 0.10 & 2.03 & 0.31 & 0.10 & 3.03 & 0.31 & 0.10 \\
 \cline{2-15}
 & \multirow{5}{2em}{Sc 2} & Baseline & -0.09 & 0.58 & 0.34 & 0.92 & 0.57 & 0.34 & 1.91 & 0.58 & 0.34 & 2.92 & 0.57 & 0.34\\
 & & PSCL1 & -0.04 & 0.36 & 0.13 & 0.97 & 0.34 & 0.12 & 1.96 & 0.36 & 0.13 & 2.97 & 0.34 & 0.12 \\
 & & PSCL2 & -0.53 & 0.31 & 0.38 & 0.47 & 0.30 & 0.37 & 1.46 & 0.30 & 0.38 & 2.46 & 0.31 & 0.38 \\
 & & PSCL3 & -0.45 & 0.35 & 0.33 & 0.55 & 0.35 & 0.32 & 1.56 & 0.35 & 0.33 & 2.55 & 0.35 & 0.32 \\
 & & PAM-HC & -0.08 & 0.28 & 0.09 & 0.94 & 0.30 & 0.09 & 1.92 & 0.28 & 0.09 & 2.93 & 0.30 & 0.09 \\
 \cline{2-15}
 & \multirow{5}{2em}{Sc 3} & Baseline & -0.09 & 0.58 & 0.34 & 0.91 & 0.56 & 0.32 & 1.91 & 0.58 & 0.34 & 2.91 & 0.56 & 0.32 \\
 & & PSCL1 & -0.09 & 0.16 & 0.03 & 0.92 & 0.16 & 0.03 & 1.91 & 0.16 & 0.03 & 2.92 & 0.16 & 0.03 \\
 & & PSCL2 & -0.49 & 0.31 & 0.33 & 0.54 & 0.29 & 0.30 & 1.51 & 0.32 & 0.34 & 2.53 & 0.29 & 0.31 \\
 & & PSCL3 & -0.15 & 0.23 & 0.08 & 0.86 & 0.22 & 0.07 & 1.85 & 0.23 & 0.08 & 2.86 & 0.22 & 0.07 \\
 & & PAM-HC & -0.03 & 0.22 & 0.05 & 0.98 & 0.22 & 0.05 & 1.97 & 0.22 & 0.05 & 2.98 & 0.22 & 0.05 \\
 \hline
 \multirow{15}{3em}{450} & \multirow{5}{2em}{Sc 1} & Baseline & -0.09 & 0.45 & 0.21 & 0.90 & 0.48 & 0.24 & 1.89 & 0.47 & 0.23 & 2.89 & 0.47 & 0.24 \\
 & & PSCL1 & 0.05 & 0.12 & 0.02 & 1.03 & 0.14 & 0.02 & 2.03 & 0.14 & 0.03 & 3.02 & 0.13 & 0.02 \\
 & & PSCL2 & 0.48 & 0.33 & 0.34 & 1.46 & 0.35 & 0.34 & 2.46 & 0.34 & 0.32 & 3.46 & 0.35 & 0.33 \\
 & & PSCL3 & 0.14 & 0.22 & 0.07 & 1.13 & 0.23 & 0.07 & 2.12 & 0.22 & 0.06 & 3.12 & 0.23 & 0.07 \\
 & & PAM-HC & 0.02 & 0.14 & 0.02 & 1.00 & 0.16 & 0.02 & 2.00 & 0.14 & 0.02 & 2.99 & 0.15 & 0.02 \\
 \cline{2-15}
 & \multirow{5}{2em}{Sc 2} & Baseline & -0.08 & 0.46 & 0.22 & 0.91 & 0.48 & 0.24 & 1.90 & 0.48 & 0.24 & 2.90 & 0.48 & 0.24 \\
 & & PSCL1 & -0.04 & 0.24 & 0.06 & 0.95 & 0.25 & 0.07 & 1.94 & 0.25 & 0.07 & 2.94 & 0.26 & 0.07  \\
 & & PSCL2 & -0.46 & 0.28 & 0.29 & 0.53 & 0.28 & 0.30 & 1.52 & 0.30 & 0.32 & 2.52 & 0.30 & 0.32 \\
 & & PSCL3 & -0.40 & 0.28 & 0.24 & 0.58 & 0.30 & 0.26 & 1.58 & 0.30 & 0.27 & 2.57 & 0.30 & 0.27 \\
 & & PAM-HC & -0.00 & 0.17 & 0.03 & 0.99 & 0.18 & 0.03 & 1.97 & 0.20 & 0.04 & 2.97 & 0.19 & 0.04 \\
 \cline{2-15}
 & \multirow{5}{2em}{Sc 3} & Baseline & -0.09 & 0.49 & 0.25 & 0.90 & 0.50 & 0.27 & 1.88 & 0.50 & 0.27 & 2.89 & 0.51 & 0.27 \\
 & & PSCL1 & -0.08 & 0.11 & 0.02 & 0.90 & 0.12 & 0.02 & 1.90 & 0.13 & 0.03 & 2.90 & 0.12 & 0.03 \\
 & & PSCL2 & -0.44 & 0.28 & 0.25 & 0.55 & 0.28 & 0.26 & 1.54 & 0.31 & 0.27 & 2.54 & 0.27 & 0.27 \\
 & & PSCL3 & -0.16 & 0.18 & 0.06 & 0.82 & 0.19 & 0.07 & 1.82 & 0.20 & 0.07 & 2.81 & 0.23 & 0.07 \\
 & & PAM-HC & 0.02 & 0.24 & 0.07 & 1.00 & 0.23 & 0.07 & 1.99 & 0.24 & 0.08 & 2.99 & 0.23 & 0.06 \\
 \hline
\end{tabular}
}
\end{center}
\end{table}

\begin{table}[!p]
\caption{Simulation results based on binary outcome for PAM-HC, baseline method, and three versions of PSCL methods. Here $\hat{\Delta}$ is the average posterior mean of the overall treatment effect across 100 simulated trials.}
\label{tab:bin_result}
\begin{center}
\resizebox{\linewidth}{!}{
\begin{tabular}{ c|c|c|ccc|ccc|ccc|ccc }
 \hline
 \multirow{2}{3em}{$N$} & \multirow{2}{2em}{Sc} & \multirow{2}{6em}{Method} & \multicolumn{3}{|c|}{True $\Delta = 0.00$*} & \multicolumn{3}{|c|}{True $\Delta = 9.42$*} & 
 \multicolumn{3}{|c|}{True $\Delta = 15.67$*} &
 \multicolumn{3}{|c}{True $\Delta = 19.52$*} \\
  & & & $\hat{\Delta}$* & SD & MSE* & $\hat{\Delta}$* & SD & MSE* & $\hat{\Delta}$* & SD & MSE* & $\hat{\Delta}$* & SD & MSE* \\
 \hline
 \multirow{15}{3em}{300} & \multirow{5}{2em}{Sc 1} & Baseline & -0.37 & 0.06 & 0.35 & 5.69 & 0.06 & 0.50 & 12.60 & 0.06 & 0.50 & 17.73 & 0.06 & 0.38 \\
 & & PSCL1 & -0.24 & 0.02  & 0.05 & 6.18 & 0.03 & 0.20 & 12.77 & 0.03 & 0.16 & 18.25 & 0.03 & 0.11 \\
 & & PSCL2 & 3.15 & 0.04 & 0.26 & 9.57 & 0.05 & 0.22 & 16.13 & 0.04 & 0.19 & 21.57 & 0.04 & 0.23 \\
 & & PSCL3 & 0.41  & 0.03 & 0.11 & 6.90 & 0.04 & 0.20 & 13.48 & 0.03 & 0.17 & 19.02 & 0.03 & 0.12  \\
 & & PAM-HC & 0.29 & 0.03 & 0.11 & 6.65 & 0.04 & 0.23 & 13.14 & 0.04 & 0.22 & 18.52 & 0.04 & 0.16 \\
 \cline{2-15}
 & \multirow{5}{2em}{Sc 2} & Baseline & -0.34 & 0.06 & 0.37 & 5.85 & 0.06 & 0.49 & 12.72 & 0.06 & 0.48 & 18.04 & 0.06 & 0.39\\
 & & PSCL1 & -4.56 & 0.04 & 0.37 & 1.76 & 0.04 & 0.76 & 8.40 & 0.04 & 0.73 & 13.94 & 0.04 & 0.48 \\
 & & PSCL2 & -4.54 & 0.04 & 0.32 & 1.78 & 0.03 & 0.68 & 8.50 & 0.03 & 0.64 & 13.84 & 0.03 & 0.42 \\
 & & PSCL3 & -6.08 & 0.04 & 0.56 & 0.25 & 0.04 & 1.01 & 6.99 & 0.04 & 0.97 & 12.42 & 0.04 & 0.70 \\
 & & PAM-HC & -0.61 & 0.03 & 0.12 & 5.67 & 0.03 & 0.26 & 12.25 & 0.03 & 0.26 & 17.69 & 0.03 & 0.16 \\
 \cline{2-15}
 & \multirow{5}{2em}{Sc 3} & Baseline & -0.48 & 0.06 & 0.34 & 6.73 & 0.06 & 0.51 & 12.50 & 0.06 & 0.46 & 17.78 & 0.06 & 0.39 \\
 & & PSCL1 & -0.34 & 0.02 & 0.06 & 6.17 & 0.03 & 0.18 & 12.68 & 0.03 & 0.16 & 18.23 & 0.03 & 0.09 \\
 & & PSCL2 & -4.01 & 0.03 & 0.25 & 2.60 & 0.03 & 0.58 & 9.01 & 0.03 & 0.55 & 14.65 & 0.03 & 0.33 \\
 & & PSCL3 & -0.50 & 0.02 & 0.06 & 6.08 & 0.03 & 0.18 & 12.51 & 0.03 & 0.17 & 18.14 & 0.03 & 0.08 \\
 & & PAM-HC & -0.30 & 0.03 & 0.08 & 6.08 & 0.03 & 0.20 & 12.49 & 0.03 & 0.20 & 17.95 & 0.03 & 0.13 \\
 \hline
 \multirow{15}{3em}{450} & \multirow{5}{2em}{Sc 1} & Baseline & -0.51 & 0.05 & 0.23 & 5.93 & 0.05 & 0.37 & 12.68 & 0.05 & 0.31 & 17.89 & 0.05 & 0.28 \\
 & & PSCL1 & -0.01 & 0.02 & 0.06 & 6.52 & 0.02 & 0.14 & 13.22 & 0.02 & 0.10 & 18.61 & 0.02 & 0.07 \\
 & & PSCL2 & 3.36 & 0.04 & 0.25 & 9.92 & 0.04 & 0.14 & 16.51 & 0.03 & 0.12 & 21.96 & 0.03 & 0.17 \\
 & & PSCL3 & 0.75 & 0.03 & 0.09 & 7.35 & 0.03 & 0.14 & 14.02 & 0.02 & 0.09 & 19.42 & 0.03 & 0.07  \\
 & & PAM-HC & 0.32 & 0.03 & 0.08 & 6.86 & 0.02 & 0.13 & 13.41 & 0.02 & 0.10 & 18.73 & 0.03 & 0.07  \\
 \cline{2-15}
 & \multirow{5}{2em}{Sc 2} & Baseline & -0.57 & 0.05 & 0.24 & 6.04 & 0.05 & 0.37 & 12.77 & 0.05 & 0.31 & 17.95 & 0.05 & 0.28\\
 & & PSCL1 & -4.33 & 0.04 & 0.31 & 2.26 & 0.03 & 0.62 & 9.06 & 0.03 & 0.56 & 14.23 & 0.04 & 0.41 \\
 & & PSCL2 & -4.30 & 0.03 & 0.28 & 2.42 & 0.03 & 0.59 & 9.17 & 0.03 & 0.53 & 14.36 & 0.03 & 0.35 \\
 & & PSCL3 & -6.20 & 0.04 & 0.52 & 0.49 & 0.04 & 0.92 & 7.36 & 0.04 & 0.85 & 12.39 & 0.04 & 0.64 \\
 & & PAM-HC & -0.22 & 0.03 & 0.08 & 6.40 & 0.03 & 0.17 & 13.08 & 0.03 & 0.14 & 18.18 & 0.03 & 0.10 \\
 \cline{2-15}
 & \multirow{5}{2em}{Sc 3} & Baseline & -0.40 & 0.05 & 0.26 & 5.70 & 0.05 & 0.41 & 12.58 & 0.05 & 0.35 & 17.85 & 0.05 & 0.31 \\
 & & PSCL1 & -0.23 & 0.02 & 0.06 & 6.18 & 0.02 & 0.15 & 12.87 & 0.02 & 0.12 & 18.32 & 0.02 & 0.06 \\
 & & PSCL2 & -3.73 & 0.03 & 0.23 & 2.65 & 0.03 & 0.54 & 9.33 & 0.02 & 0.47 & 14.82 & 0.03 & 0.29 \\
 & & PSCL3 & -0.43 & 0.02 & 0.05 & 5.96 & 0.02 & 0.18 & 12.64 & 0.02 & 0.14 & 18.06 & 0.02 & 0.08 \\
 & & PAM-HC & 0.16 & 0.03 & 0.10 & 6.49 & 0.03 & 0.20 & 13.18 & 0.02 & 0.15 & 18.49 & 0.03 & 0.11 \\
 \hline
 \multicolumn{10}{l}{* True $\Delta$, $\hat{\Delta}$, and MSE values times 100.} \\
\end{tabular}
}
\end{center}
\end{table}

\end{document}